\documentclass[aps,prd,preprint,showpacs,nofootinbib,11pt]{revtex4-1}
\usepackage{latexsym,bm,amsmath,amssymb}

\begin{document}

\title{Stationary Lifshitz black holes of $R^2$-corrected gravity theory}

\author{{\"O}zg{\"u}r Sar{\i}o\u{g}lu}
\email{sarioglu@metu.edu.tr}
\affiliation{Department of Physics, Faculty of Arts and  Sciences,\\
             Middle East Technical University, 06800, Ankara, Turkey}

\date{\today}

\begin{abstract}
In this short note, I present a generalization of a set of \emph{static} $D$-dimensional ($D \geq 3$) 
Lifshitz black holes, which are solutions of the gravitational model obtained by amending the cosmological 
Einstein theory with the addition of only the curvature-scalar-squared term and that are described by two 
parameters, to a more general class of exact, analytic solutions that involves an additional parameter 
which now renders them \emph{stationary}. In the special $D=3$ and the dynamical exponent $z=1$ case, 
the parameters can be adjusted so that the solution becomes identical to the celebrated BTZ black hole metric.
\end{abstract}


\maketitle

In recent years a flurry of activity has been seen on the extensions of the celebrated AdS/CFT correspondence
to diverse areas of physics, such as non-relativistic, and more specifically, condensed matter systems (see
\cite{son,bala1} and the references that cite these e.g. \cite{abm,hart} for a sample). Among works in this 
vein, \cite{kach} is worthy of mentioning, since it was there that the Lifshitz spacetimes, which have ever 
since enjoyed a growing interest as gravity duals of theories with nontrivial scaling properties [(and with 
fewer symmetries) compared to their cousins which are conformal and dual to AdS spacetimes], were first 
introduced into the literature. On the other hand, black hole solutions which asymptote to these Lifshitz 
spacetimes, and thus which are simply referred to as ``Lifshitz black holes'', are needed when describing 
finite temperature aspects of these non-relativistic systems. However, there aren't that many exact analytic 
Lifshitz black holes around, and among those that have been found so far (see e.g. 
\cite{bala2,peet,giri,mann,bdtz,pang} and the references therein), there are only \emph{static} ones and, 
perhaps surprisingly, there are no known \emph{stationary} ones at all. This is the main motivation of the 
present work. Here, I find a class of stationary $D$-dimensional exact analytic spacetimes, which are solutions 
of the $R^2$-corrected gravity theory, and discuss the conditions which allow these to be interpreted as Lifshitz 
black holes.

The action of the gravity theory I consider is
\begin{equation}
 I = \int \, d^{D}x \, \sqrt{-g} \, \Big( R + 2 \Lambda + \alpha R^{2} \Big) \,, \label{act}
\end{equation}
where $\Lambda$ is the cosmological constant, and the field equations that follow from the variation of
(\ref{act}) are
\begin{equation}
R_{ab} - \frac{1}{2} g_{ab} R - \Lambda g_{ab} + \alpha \Big( 2 R R_{ab} - 2 \nabla_{a} \nabla_{b} R 
 + g_{ab} (2 \, \square R - \frac{1}{2} R^2) \Big) = 0 \,, \label{feq}
\end{equation}
with \( \square \equiv \nabla_{c} \nabla^{c} \).

Aside from these, I also demand the modified scaling transformations
\[ t \mapsto \lambda^{z} t \,, \quad \vec{x} \mapsto \lambda \vec{x} \,, \quad r \mapsto r/\lambda \,, \]
where $\vec{x}$ denotes a $(D-2)$-dimensional vector, that are respected by the \emph{static} Lifshitz 
spacetimes to hold as usual. I also want to keep the invariance under time and space translations, and
spatial rotations intact; however instead of separately asking for spatial parity (P) and time reversal (T)
invariance, I go for the weaker PT invariance.\footnote{To be precise, I also do not include the 
transformation \( r \mapsto -r \) to the allowed P transformations, or the translation \( r \mapsto r - r_{0} \)
to the spatial translations, in accordance with the nomenclature on the Lifshitz spacetimes in the literature. 
Thus, e.g. I think of PT as PT: \( (t,r,\phi,\vec{x}) \mapsto (-t,r,-\phi,-\vec{x}) \) throughout.} These 
assumptions lead to the following \emph{stationary} spacetimes
\begin{equation}
 ds^2 = - \frac{r^{2 z}}{\ell^{2 z}} \, dt^2 + 2 \omega \, \frac{r^{z+1}}{\ell^{z+1}} \, dt \, d\phi 
  + \frac{r^2}{\ell^2} \, d\phi^2 + \frac{\ell^2}{r^2} \, dr^2 + \frac{r^2}{\ell^2} \, d\vec{x}^2 \,, 
 \label{lifsp}
\end{equation}
where \( 0 \leq r < \infty \) and \( d\vec{x}^2 \equiv \sum_{i=1}^{D-3} dx_{i}^{2} \), which obviously 
reduce to the usual static Lifshitz spacetimes when the `rotation parameter' $\omega$ is set to zero. For later 
convenience, I have chosen \( x_{D-2} \equiv \phi \) here and separated it from the remaining 
\( x_{i} \; (1 \leq i \leq D-3) \). Clearly $\ell>0$ sets the length scale in this geometry, and the 
parameter $\omega$, just like the `dynamical exponent' $z$, is dimensionless.

One finds that (\ref{lifsp}) is a solution to (\ref{feq}) for generic values of the parameters $z$
and $\omega$ in any $D \geq 3$, provided that the coupling constant $\alpha$ and the cosmological
constant $\Lambda$ are chosen as
\begin{equation}
 \alpha = \frac{1}{8 \Lambda} \,, \quad
 \Lambda = \frac{2 D^2 + 3 (z - 1)^2 + 2 D (2 z - 3)}{8 \ell^{2}} + \frac{(z - 1)^2}{8 \ell^{2} (1 + \omega^2)} \,.
 \label{alpLam}
\end{equation}
However, there is even more to the story: Provided that the coupling constants in the action (\ref{act}) are 
chosen precisely as in (\ref{alpLam}), the following metric 
\begin{eqnarray}
 ds^2 & = & - \frac{r^{2 z}}{\ell^{2 z}} \, h(r) \, dt^2 + \frac{r^2}{\ell^2} \Big( d\phi + \omega \, \frac{\ell^2}{r^2} \, dt \Big)^2 
 + \frac{\ell^2}{r^2} \, \frac{dr^2}{h(r)} + \frac{r^2}{\ell^2} \, d\vec{x}^2 \,, \; \mbox{where} \label{met} \\
 h(r) & \equiv & c + k \, \frac{\ell^{2(1+z)}}{r^{2(1+z)}} + M^{-} \, \frac{\ell^{p_{-}}}{r^{p_{-}}} 
                 + M^{+} \, \frac{\ell^{p_{+}}}{r^{p_{+}}} \,, \; \mbox{with} \label{hfun} \\
 c & \equiv & \frac{4 \, \ell^2 \, \Lambda}{2 z^2 + (D-2)(2 z + D - 1)} \,, \;\; k \equiv \frac{2 \, \omega^2}{D^2 -7 D + 14 - 2 z (D-3)} 
                 \,, \; \mbox{and} \label{candk} \\ 
 p_{\pm} & \equiv & \frac{1}{2} \Big( 3 z +2 (D-2) \pm \sqrt{z^2 + 4(D-2)(z-1)} \Big) \,, 
\label{ppm}
\end{eqnarray}
turns out to be a solution of the field equations (\ref{feq}) for any dimension $D \geq 3$. Note that the
coefficients $c$ and $k$ are completely determined by $z$ and $\omega$, whereas the integration constants 
$M^{\pm}$ are left as free parameters. 

Before proceeding any further, it must be stated that the static version of the metrics (\ref{lifsp}) and
(\ref{met}) [ i.e. those with vanishing $\omega$ for which $c=1$, $k=0$, the relation (\ref{alpLam}) for the 
constants in the action (\ref{act}) and the metric function $h(r)$ in (\ref{hfun}) are simplified accordingly]
were first presented in Sec. 2 of \cite{giri}. However, with the turning on of the parameter $\omega$,
the stationary metrics (\ref{lifsp}) and (\ref{met}) (with the accompanying equations (\ref{alpLam}) and
(\ref{hfun})-(\ref{ppm}), respectively) are clearly more general than the solutions given in \cite{giri}.

To note a rather appealing feature of these solutions and to give an interesting example, let me quickly
consider the conformal limit $z=1$ and, for simplicity, set $D=3$. One finds that the metric (\ref{met}) 
becomes identical to the BTZ metric \cite{btz} when one sets \( M^{+} = 0, M^{-} = - M < 0 \) and 
\( \omega = - j/2 \):
\[ ds^2 = \Big( M - \frac{r^2}{\ell^2} \Big) \, dt^2 - j \, dt \, d\phi + \frac{r^2}{\ell^2} \, d\phi^2
 + \frac{dr^2}{- M + \frac{r^2}{\ell^2} + \frac{j^2 \ell^2}{4 r^2}} \,, \]
which can be further brought to the canonical form when one takes \( \phi = \theta \ell \) first
and later sets \( J = j \ell \).

Let me also note that the curvature scalars of the metrics (\ref{lifsp}) and (\ref{met}) are both given by
\( R = - 4 \Lambda \) precisely. As discussed earlier in \cite{giri} and \cite{biz2}, this allows for the
casting of the action (\ref{act}) into the form
\[  I = \frac{1}{8 \Lambda} \int \, d^{D}x \, \sqrt{-g} \, (R + 4 \Lambda)^{2} \,, \]
and this theory cannot be mapped into a scalar-tensor theory by a conformal transformation of the metric.
Thus one is really dealing with an authentic gravity theory here \cite{giri}.

To have $h(r)$ real, it follows from the expression of $p_{\pm}$ that 
\( z^2 + 4 (D-2)(z-1) > 0 \), which further implies that\footnote{In the discussion
that follows, the points $z = z_{-}$ and $z = z_{+}$ are excluded from the allowed range of the
dynamical exponent $z$ for a good reason: At these critical values, the metric function $h(r)$
given in (\ref{hfun}) is no longer valid and needs to be modified. This is discussed in detail for
the $z = z_{+}$ case in the paragraph containing Eqs. (\ref{smet})-(\ref{ssigp}) below.}
\[ z < z_{-} \equiv 4 - 2 D - 2 \sqrt{(D-1)(D-2)} < 0 \quad \mbox{or} \quad
   z > z_{+} \equiv 4 - 2 D + 2 \sqrt{(D-1)(D-2)} > 0  \,. \]
However, if one is also to demand that the metric (\ref{met}) describes a black hole solution,
then one can consider the branch \( z > z_{+} > 0 \) since then one can show, for the powers of $r$
in the metric function $h(r)$, that \( p_{+} \geq p_{-} \geq z_{+} > 0 \) and \( 2 (z+1) > 0 \). Hence 
for \( z > z_{+} \), the solution (\ref{met}) represents a black hole with an event horizon located 
at $r = r_{+}$, that is found by the largest positive real root of $h(r)$ i.e. $h(r_{+}) = 0$, provided the 
remaining parameters $\omega$ (thus $c$ and $k$) and $M^{\pm}$ also play along appropriately. 

Obviously, a general analysis with generic values of $z$ and $\omega$ given a specific $D$ is a 
complicated problem. However, it should be possible in principle to ``exorcise'' other exotic solutions 
such as extremal black holes given the generic solution (\ref{met}) by ``playing around'' with the 
parameters $M^{\pm}$ and/or $z$ and $\omega$. An example in this vein is obtained when one 
considers the special case where the critical value for the dynamical exponent $z$ is reached and
one sets $z = z_{+}$ exactly. The solution now reads
\begin{eqnarray}
 ds^2 & = & - \frac{r^{2 z_{+}}}{\ell^{2 z_{+}}} \, f(r) \, dt^2 
 + \frac{r^2}{\ell^2} \Big( d\phi + \omega \, \frac{\ell^2}{r^2} \, dt \Big)^2 
 + \frac{\ell^2}{r^2} \, \frac{dr^2}{f(r)} + \frac{r^2}{\ell^2} \, d\vec{x}^2 \,, \; \mbox{where} \label{smet} \\
 f(r) & \equiv & c_{+} + k_{+} \, \frac{\ell^{2(1+z_{+})}}{r^{2(1+z_{+})}}  
                 + \frac{\ell^{p}}{r^{p}} \Big( M_{1} + M_{2} \ln{(r/\ell)} \Big) \,, \; \mbox{with} \label{ffun} \\
 s & \equiv & c_{+} \, \Big{\vert}_{z = z_{+}}  \,, \;\; k_{+} \equiv k \, \Big{\vert}_{z = z_{+}} \,, 
 \;\; \mbox{and} \;\; p \equiv p_{+} \, \Big{\vert}_{z = z_{+}} \,, \label{ssigp} 
\end{eqnarray}
and $M_{1}$ and $M_{2}$ are the new free parameters. Note that this is very similar in form to the 
previous one (\ref{met})-(\ref{ppm}), but now the metric function $f(r)$ asymptotes to a constant in 
a slower fashion when one approaches the boundary at $r \to \infty$.

In the discussion above, I had to set \( z <  z_{-} < 0 \) or \( z > z_{+} > 0 \) so that the metric 
function $h(r)$ in (\ref{hfun}) is real in the first place. One cannot help but wonder whether this 
really means that the region \( z \in (z_{-},z_{+}) \) is completely `forbidden' for the dynamical 
exponent $z$? Working out the field equations for the special value $z=0$ indicates that this is 
certainly not the case! For $z=0$, the solution is
\begin{eqnarray}
 ds^2 & = & - g(r) \, dt^2 + \frac{r^2}{\ell^2} \Big( d\phi + \omega \, \frac{\ell^2}{r^2} \, dt \Big)^2 
 + \frac{\ell^2}{r^2} \, \frac{dr^2}{g(r)} + \frac{r^2}{\ell^2} \, d\vec{x}^2 \,, \; \mbox{where} \label{0met} \\
 g(r) & \equiv & c_{0} + k_{0} \, \frac{\ell^{2}}{r^{2}}  
                 + \frac{\ell^{D-2}}{r^{D-2}} \Big( M_{c} \, \cos{\big( \sqrt{D-2} \, \ln{(r/\ell)} \big)} 
                 + M_{s} \, \sin{\big( \sqrt{D-2} \, \ln{(r/\ell)} \big)} \Big) \,, \; \mbox{with} \label{gfun} \\
 c_{0} & \equiv & c \, \Big{\vert}_{z = 0} \;\; \mbox{and} \;\; k_{0} \equiv k \, \Big{\vert}_{z = 0} \,, \label{0ck} 
\end{eqnarray}
and $M_{c}$ and $M_{s}$ are the two integration constants.

It is clearly of paramount importance to compute the conserved charges and to study the thermodynamical
properties of the solutions presented here to better understand their geometrical and physical aspects. This
should also be relevant for their dual condensed matter systems. In \cite{biz2}, the solutions 
(\ref{met})-(\ref{ppm}) with vanishing $\omega$ were studied using the background Killing charge
definition developed in \cite{biz1}, and they were shown to have both their energy $E=0$ and their
entropy $S=0$. One would expect this result to remain intact, i.e. that both $E=0$ and $S=0$,
which, taken together in the first law of black hole thermodynamics, also implies that the angular 
momentum $J$ vanishes as well: $J=0$. This should make all the more sense when one considers the
remark on the vanishing of the action $I$ that follows from \( R = - 4 \Lambda \) at the first place.

As a concrete and illustrative example along these lines, let me now concentrate on the $D=4$, $z=3/2$
case. This particular choice is sort of motivated by historical reasons: The corresponding solution
(\ref{met})-(\ref{ppm}) together with $M^{+} = 0$ and $\omega = 0$ was the first ever member of the
black hole solutions generalized here and was first given in \cite{cinli}. In this case, the cosmological
constant $\Lambda$ and the metric function $h(r)$ in (\ref{hfun}) read
\[ \Lambda = \frac{132 + 131 \omega^2}{32 \ell^2 (1 +  \omega^2)} \;\; \mbox{and} \;\;
 h(r) = \frac{132 + 131 \omega^2}{132 (1 + \omega^2)} - 2 \omega^2 \, \frac{\ell^{5}}{r^{5}} 
 + M^{-} \, \frac{\ell^{3}}{r^{3}} + M^{+} \, \frac{\ell^{11/2}}{r^{11/2}} \,, \]
respectively. Now the background Killing vector that yields the energy is 
\( \bar{\xi}^{a} = - (\partial/\partial t)^{a} \), and the one that gives the angular momentum is
\( \bar{\varsigma}^{a} = (\partial/\partial \phi)^{a} \), in the notation of \cite{biz2}. However,
there seems to be a nontrivial ambiguity in the choice of the relevant background. If one naively,
but somehow more intuitively, chooses the background to be the usual \emph{static} Lifshitz 
spacetime with $z=3/2$
\begin{equation}
 ds^2 = - \frac{r^{3}}{\ell^{3}} \, dt^2 + \frac{r^2}{\ell^2} \, d\phi^2 + \frac{\ell^2}{r^2} \, dr^2 
  + \frac{r^2}{\ell^2} \, dx^2 \,, \label{back}
\end{equation}
which was the background that was originally employed in \cite{biz2} anyway, one
surprisingly and unexpectedly finds the divergent, unphysical results $E \to \infty$ and $J \to \infty$.
The background that leads to the aforementioned predictions, i.e. $E=0$ and $J=0$ result, is obtained 
when one sets  $M^{\pm} \to 0$ but leaves $\omega$ on, i.e. \( \omega \neq 0 \), in $h(r)$ above and 
uses this `undressed' $h(r)$ in (\ref{met}) with $z=3/2$. 

Obviously, the former (\ref{back}) and the latter backgrounds are quite different from each other both
locally and structurally. This difference may partially account for the discrepancy between the two 
energy and angular momentum computations, but I believe this is not the proper place for a discussion 
on this point, and that the issue itself is beyond the scope of the present work. Obviously, a more detailed 
and robust calculation of the conserved charges using other gravitational charge definitions would be of 
great value in studying the thermodynamics of the solutions presented here. Other open problems worthy 
of further study involve an analysis regarding the stability of these solutions and a detailed discussion, 
similar to the one in \cite{kach}, on the implications of these geometries on the condensed matter systems 
that they are related to by holographic techniques.

\begin{acknowledgments}
This work is partially supported by the Scientific and Technological Research Council of Turkey (T{\"U}B\.{I}TAK).
\end{acknowledgments}

\end{document}